\documentclass[a4,oneside,english,reqno,12pt]{amsart}
\usepackage{geometry}
\usepackage{amssymb,amsmath,amsthm,amsfonts,mathtools,esint,dsfont,bbm,yhmath,breqn}
\usepackage{multirow}
\usepackage{tabularx}
\usepackage{babel}
\usepackage{subcaption}
\usepackage{tikz}
\usepackage{graphicx}
\usepackage{float}
\usepackage{enumitem}
\usepackage{appendix}

\usepackage{hyperref}
\usepackage{bookmark}
\usepackage{cleveref}
\usepackage{crossreftools}
\crefname{subsection}{subsection}{subsections}
\usepackage[foot]{amsaddr}
\usepackage{orcidlink}
\makeatletter
\theoremstyle{plain}

\theoremstyle{remark}

\providecommand{\theoremname}{Theorem}

\allowdisplaybreaks
\begin{document}
	
	\title[VORTEX PATTERNS OF A 2D ROTATING BEC]{VORTEX PATTERNS OF A TWO-DIMENSIONAL BOSE--EINSTEIN CONDENSATE AT THE ALMOST CRITICAL ROTATION SPEED}
	
	\author[B.-D. Le]{Bao-Duy Le$^{1,3}$}
	\address{$^1$Faculty of Physics and Engineering Physics, University of Science, Ho Chi Minh City, Vietnam.}
	\email{\href{mailto:24130012@student.hcmus.edu.vn}{24130012@student.hcmus.edu.vn}}
	
	\author[D.-T. Nguyen]{Dinh-Thi Nguyen$^{2,3}$}
	\address{$^2$Faculty of Mathematics and Computer Science, University of Science, Ho Chi Minh City, Vietnam.}
	\address{$^3$Vietnam National University, Ho Chi Minh City, Vietnam.}
	\email{\href{mailto:ndthi@hcmus.edu.vn}{ndthi@hcmus.edu.vn}}
	
	\begin{abstract}
		We study vortex patterns of a two-dimensional Bose--Einstein condensate rotating close to the centrifugal limit, treating the two signs of the contact interaction with the method each requires: for repulsion, a GPU-accelerated variational minimization with exact projection onto the Lowest Landau Level (LLL); for attraction, imaginary-time evolution of the full real-space Gross--Pitaevskii (GP) equation. For repulsive interactions, our approach reproduces Abrikosov vortex lattices, achieving quantitative alignment with Thomas--Fermi theory and recovering the Abrikosov constant $e^{\rm Ab}(1)\approx1.1596$, in close analogy with the vortex ordering of type-II superconductors. In the attractive regime, the rotating ground state carries no vortex lattice at any rotation frequency: the cloud contracts and collapses as $G\to -G_*$, where $G_*=\|Q\|_{L^2}^2\simeq11.70$, essentially independently of rotation. The only stationary vortex states we find are giant vortices, trapped vortex (Townes) solitons whose charge-$m$ sector collapses as $G\to -\|Q_m\|_{L^2}^2$ ($\|Q_m\|_{L^2}^2\simeq11.70,\,48.3,\,89.8,\,132.4$ for $m=0,1,2,3$), fixed by an equivariant Gagliardo--Nirenberg inequality and confirmed numerically up to the critical region. These findings provide a numerical benchmark for vortex formation and collapse in rotating two-dimensional quantum gases.
	\end{abstract}
	
	
	\maketitle
	
	
	\section{Introduction}
	
	Since the groundbreaking experimental realization of Bose--Einstein condensation (BEC) in dilute alkali vapors in 1995 \cite{AEMWC1995}, the study of ultra-cold quantum many-body systems has undergone a remarkable evolution. A BEC forms when a significant fraction of bosons coherently occupies the lowest quantum state, giving rise to striking phenomena such as superfluidity, the formation of quantized vortices, and macroscopic quantum coherence. While early theoretical models focused on non-interacting Bose gases, subsequent research has illuminated the pivotal role of inter-particle interactions in shaping the condensate's properties, dynamics, and stability \cite{DGPS1999}.
	
	\subsection{Many-body Hamiltonian}
	
	The many-body Hamiltonian provides the foundational framework for describing interacting bosonic systems. For a system of $N$ identical bosons confined in a two-dimensional harmonic trap rotating at angular frequency $\Omega$, the Hamiltonian is expressed as (in the units of Planck constant and particle mass)
	\begin{equation}
		H_N = \sum_{j=1}^N \left( -\frac{1}{2} \Delta_{x_j} - \Omega L_{z_{j}} + \frac{1}{2} |{x_j}|^2 \right) + \sum_{i < j} \kappa \delta_{0}(x_i - x_j).
		\label{eq:Hamiltonian}
	\end{equation}
	Here $L_{z_j}$ denotes the angular momentum operator, defined by $L_{z_j} = -\mathrm{i} x_j^{\perp} \cdot \nabla_{x_j}$, with $x_j \in \mathbb{R}^2$ the position of the $j^{\rm th}$ particle and $z_j \in \mathbb{C}$ its associated complex coordinate. Furthermore, $\kappa \in \mathbb{R}$ denotes the interaction strength, with positive values indicating repulsion and negative values indicating attraction. The delta-function interaction approximates the short-range two-body potential typical of dilute gases \cite{LSPS2005}.
	
	\subsection{Interaction parameter}
	The sign of $\kappa$ dictates the stability of rotating condensates. Repulsive interactions ($\kappa>0$) stabilize the system and give rise to quantized vortex lattices, analogous to Abrikosov patterns in superconductors \cite{NR2022}. Attractive interactions ($\kappa<0$), by contrast, lead to collapse beyond the sharp Gagliardo--Nirenberg (GN) threshold \eqref{eq:GN}, where the blow-up profile is given by the unique positive radial ground state \cite{DNR2023}. In the rapid-rotation limit $\Omega\to 1$, dynamics are confined to the Lowest Landau Level (LLL), yielding Thomas--Fermi–like (TF) vortex crystals for repulsive gases, while attractive gases display vortices only transiently before destabilization \cite{weizhu2014, ABD2005}.
	
	We treat the two regimes with complementary numerical methods, and make no claim of a single unified framework. For $\kappa>0$, constrained minimization of the Gross--Pitaevskii (GP) functional within an exact LLL basis, accelerated by automatic differentiation on GPUs, reproduces the Abrikosov energy $e^{\rm Ab}(1)$~\eqref{eq:eAb_def} with systematic convergence. For $\kappa<0$, where the relevant state is sharply peaked and an LLL truncation would discard precisely the high-momentum components that build the collapsing core, we instead integrate the full real-space GP equation in imaginary time; this captures the vortex-free contraction of the ground state and the GN-governed collapse at stronger coupling, and identifies giant vortices -- trapped vortex (Townes) solitons -- as the only stationary vortex states, each with its own charge-dependent collapse threshold. Together, these results offer a numerical validation of two cornerstone regimes in rapidly rotating 2D BECs: vortex crystallization and critical blow-up.
	
	\section{Theoretical Framework}
	\subsection{Mean-field Approximation}
	Consider a system of $N$ identical bosons in $\mathbb{R}^2$, confined by a harmonic potential and rotating with angular velocity $\Omega \in [0,1)$. The associated quantum Hamiltonian in the rotating frame is defined in \eqref{eq:Hamiltonian}. In the mean-field limit $N\to\infty$, the coupling is scaled as $\kappa=\kappa_N$ so that $2\kappa_NN\to G$ is fixed. Under this scaling, the ground state exhibits complete BEC (see \cite{LSPS2005,LNR2016}): the many-body wavefunction concentrates on product states $\Psi_N(x_1, \ldots, x_N) \approx \prod_{j=1}^N \psi(x_j)$, and the energy per particle converges to the nonlinear functional with $\|\psi\|_{L^2}^2 = 1$,
	\begin{equation}
		\mathcal{E}^{\mathrm{GP}}_{\Omega, G}[\psi]
		= \int_{\mathbb{R}^2} \left[
		\frac{1}{2} \big|(-\mathrm{i}\nabla - \Omega x^\perp)\psi(x)\big|^2
		+ \frac{G}{4} |\psi(x)|^4 + \frac{1 - \Omega^2}{2} |x|^2 |\psi(x)|^2
		\right] \mathrm{d}x.
		\label{eq:gp_energy_rbec}
	\end{equation}
	The effective interaction strength is $G=2\kappa_NN$ (or its limiting value), and its sign determines whether the interaction is repulsive ($G>0$) or attractive ($G<0$).
	
	In the rotating frame with angular velocity $\Omega$, the GP functional acquires the contribution $-\Omega L_{z_{j}}$. Completing the square in the kinetic and trapping terms yields
	\begin{equation}
		V^{\mathrm{eff}}_{\Omega}(x) =  \frac{(1-\Omega^2)|x|^2}{2}.
		\label{eq:Veff}
	\end{equation}
	This effective potential captures the competition between harmonic confinement and centrifugal forces. As $\Omega \to 1$, the trap weakens and the condensate enters the LLL, where dynamics resemble those of charged particles in a magnetic field \cite{ABD2005, ABN2006}. Repulsive gases then form ordered vortex lattices or giant-vortex states, whereas attractive gases lose stabilization and collapse. Thus, the balance of $V^{\mathrm{eff}}_{\Omega}$ and rotation governs the density profile, vortex geometry, and onset of instabilities.

	\subsection{Lowest Landau Level regime}
	In the fast-rotation limit $\Omega \to 1$, centrifugal and trapping terms nearly cancel, confining the condensate to the LLL \cite{ABN2006}. Wave functions then take the analytic form
	\begin{equation}
		\psi(z) = f(z) e^{-\frac{\Omega |z|^2}{2}}, \quad z=x+{\rm i}y,
		\label{eq:LLL}
	\end{equation}
	where $f(z)$ is an entire analytic (holomorphic) function. Moreover, we can define $f(z)$ by an ansatz \cite{ABN2006_vortex} for the LLL in the form:
	\begin{equation}
		\psi(z) = c\prod_{j=1}^J (z -z_{j}) e^{-\frac{\Omega|z|^2}{2}},
		\label{eq:seed}
	\end{equation}
	where $c$ is fixed by the $L^2$ normalization $\|\psi\|_{L^2}=1$ and $z_1,...,z_J$ are the locations of the zeros of the analytic function associated with $\psi$. Vortices are encoded in the zeros of $f$, forming triangular lattices with infinitely many zeros, including “invisible’’ ones outside the TF region \cite{GGT2018}. The projected LLL energy functional reads
	\begin{equation}
		\mathcal{E}^{\rm LLL}_{\Omega,G}[\psi] = \Omega + \int_{\mathbb{R}^2} \left[\frac{G}{4}|\psi(x)|^4 + \frac{1-\Omega^2}{2}|x|^2|\psi(x)|^2 \right] \mathrm{d}x.
		\label{eq:E_gp}
	\end{equation}
	
	\subsection{Thomas--Fermi theory}
	In the repulsive regime of trapped dilute BECs, the TF approximation provides an accurate macroscopic density profile when kinetic energy is negligible \cite{DGPS1999}. Under rapid rotation, the condensate is confined to the LLL and the GP functional reduces to an effective TF form \cite{NR2022,NR2025}:
	\begin{equation}
		\mathcal{E}^{\mathrm{LLL}}_{\Omega,G}[\psi] - \Omega
		\underset{\Omega \to 1}{\approx}
		\mathcal{E}^{\mathrm{\rm TF}}_{\Omega,G}[\rho]
		:= \int_{\mathbb{R}^2}\left[\frac{e^{\mathrm{Ab}}(1)G}{4}\rho(x)^2
		+ \frac{1-\Omega^2}{2}|x|^2\rho(x)\right] \mathrm{d}x.
		\label{eq:TF_LLL}
	\end{equation}
	The Abrikosov constant $e^{\mathrm{Ab}}(1)$ characterizes the minimal quartic interaction energy in the LLL \cite{ABN2006}, i.e.,
	\begin{equation}
		e^{\mathrm{Ab}}(1)
		= \lim_{L\to\infty}
		\inf_{\substack{\psi \in \mathcal{LLL}_L \\ \|\psi\|_{L^2(Q_L)}=1}}
		\frac{1}{|Q_L|}\int_{Q_L} |\psi(x)|^4 \, {\rm d}x,
		\label{eq:eAb_def}
	\end{equation}
	where $\mathcal{LLL}_L$ denotes the finite-dimensional LLL restricted to a square cell $Q_L$ of side length $L$, with magnetic-periodic boundary conditions and flux quantization
	\[
		|Q_L|=L^2\in 2\pi\mathbb{N}.
	\]
	Equivalently, this is the thermodynamic limit, taken along such admissible cells, of $\langle|\psi|^4\rangle/\langle|\psi|^2\rangle^2$. For the non-rotating case ($\Omega=0$, $G=1$), neglecting the kinetic energy in \eqref{eq:gp_energy_rbec} gives
	\begin{equation}
		\rho^{\mathrm{TF}}_{0,1}(x) = \left[ \lambda^{\mathrm{TF}}_{0,1} - |x|^2 \right]_+,
		\label{eq:TF-rotating}
	\end{equation}
	where $\lambda^{\mathrm{TF}}_{0,1}$ is the support parameter. The normalization condition gives
	\begin{equation}
		1=2\pi\int_0^{\sqrt{\lambda^{\mathrm{TF}}_{0,1}}}\left(\lambda^{\mathrm{TF}}_{0,1}-r^2\right)r\,{\rm d}r
		=\frac{\pi}{2}\left(\lambda^{\mathrm{TF}}_{0,1}\right)^2,
		\qquad
		\lambda^{\mathrm{TF}}_{0,1}=\sqrt{\frac{2}{\pi}}.
	\end{equation}
	The Euler--Lagrange equation 
	$$
	\frac{\rho^{\mathrm{TF}}_{0,1}}{2}+\frac{|x|^2}{2}=\mu^{\mathrm{TF}}_{0,1}
	$$
	then gives
	\begin{equation}
		\mu^{\mathrm{TF}}_{0,1}=\frac{\lambda^{\mathrm{TF}}_{0,1}}{2}
		=\frac{1}{\sqrt{2\pi}}.
	\end{equation}
	Thus the support parameter and the chemical potential are distinct even in the non-rotating case. This is the two-dimensional, unit-mass version of the usual inverted-parabola TF law~\cite{FS2001,DGPS1999}. For $\Omega \to 1$, the LLL-projected TF profile associated with \eqref{eq:TF_LLL} takes the form~\cite{NR2022}
	\begin{equation}
		\rho_{\Omega,G}^{\mathrm{TF}}(x) = \frac{1-\Omega^2}{e^{\mathrm{Ab}}(1)G}
		\left(\lambda_{\Omega,G}^{\mathrm{TF}} - |x|^2\right)_+,
		\label{eq:p_TF}
	\end{equation}
	where $\lambda_{\Omega,G}^{\mathrm{TF}} >0$ is an effective TF support parameter, introduced to distinguish it from the chemical potential. In particular, it determines the cloud radius through $R_{\Omega, G}^{\mathrm{TF}}=\sqrt{\lambda_{\Omega,G}^{\mathrm{TF}}}$. The normalization condition $\int_{\mathbb{R}^2}\rho_{\Omega,G}^{\mathrm{TF}}(x)\,\mathrm{d}x=1$ gives
	\begin{equation}
		1=\frac{\pi(1-\Omega^2)}{2e^{\mathrm{Ab}}(1)G}\left(\lambda_{\Omega,G}^{\mathrm{TF}}\right)^2,
		\qquad
		\lambda_{\Omega,G}^{\mathrm{TF}} = \left[\frac{2e^{\mathrm{Ab}}(1)G}{\pi(1-\Omega^2)}\right]^{1/2}.
		\label{eq:mu_TF}
	\end{equation}
	To identify the chemical potential, let $\mu_{\Omega,G}^{\mathrm{TF}}$ denote the Lagrange multiplier for the unit-mass constraint in \eqref{eq:TF_LLL}. On the support of the minimizer, the Euler--Lagrange equation reads
	\begin{equation}
		\frac{e^{\mathrm{Ab}}(1)G}{2}\rho_{\Omega,G}^{\mathrm{TF}}(x)
		+\frac{1-\Omega^2}{2}|x|^2
		=\mu_{\Omega,G}^{\mathrm{TF}}.
	\end{equation}
	Comparison with \eqref{eq:p_TF} therefore yields
	\begin{equation}
		\mu_{\Omega,G}^{\mathrm{TF}}
		=\frac{1-\Omega^2}{2}\lambda_{\Omega,G}^{\mathrm{TF}}
		=\left[\frac{e^{\mathrm{Ab}}(1)G(1-\Omega^2)}{2\pi}\right]^{1/2}.
	\end{equation}
	Thus $\mu_{\Omega,G}^{\mathrm{TF}}$ is the TF chemical potential (equivalently, the mass-constraint Lagrange multiplier), whereas $\lambda_{\Omega,G}^{\mathrm{TF}}$ is the associated support parameter. When the coupling is expressed in terms of the particle number, this Lagrange-multiplier characterization agrees with the usual thermodynamic definition of the chemical potential.
	
	\subsection{Cubic GP and blow-up for attractive interactions}
	\label{sb:cubic_GP}
	
	For attractive interactions ($G<0$), the 2D condensate is governed at the mean-field level by the focusing cubic GP equation
	\begin{equation}
		\mathrm{i}\partial_t \psi = \left[-\frac{1}{2}\Delta + V^{\mathrm{eff}}_{\Omega=0} - \Omega L_{z} + \frac{G}{2}|\psi|^2 \right]\psi,
		\label{eq:gp}
	\end{equation}
	with $\psi \in H^{1}(\mathbb{R}^2) = \left\{ \psi \in L^2(\mathbb{R}^2) \;\big|\; \nabla \psi \in L^2(\mathbb{R}^2) \right\}$; the nonlinear coefficient is $G/2$ because the energy \eqref{eq:gp_energy_rbec} carries $G/4$. In this regime, solutions may blow up in finite time, corresponding to macroscopic collapse \cite{thi2020,LNR2018,BDN2023}. Seeking stationary states $\psi(x,t)=e^{-\mathrm{i}\mu t}u(x)$ leads to the nonlinear elliptic problem
	\begin{equation}
		-\frac{1}{2}\Delta u + V^{\mathrm{eff}}_{\Omega=0}u - \Omega L_z u + \frac{G}{2}|u|^2u = \mu u,
	\end{equation}
	which admits solutions, where $\mu$ is the chemical potential. In the $L^2$-critical case, collapse profiles converge (after rescaling) to the unique positive radially symmetric ground state $Q\in H^1(\mathbb{R}^2)$ solving
	\begin{equation}
		-\Delta Q + Q - Q^3 = 0,
		\label{eq:elliptic_Q}
	\end{equation}
	Writing $q:=Q/\|Q\|_{L^2}$ for the unit-mass Townes profile, the normalized collapsing state has the universal asymptotics
	\begin{equation}
		u(x)\;\sim\;\frac{1}{\lambda_G}q\!\left(\frac{x}{\lambda_G}\right)
		=\frac{1}{\lambda_G\|Q\|_{L^2}}Q\!\left(\frac{x}{\lambda_G}\right),
		\label{eq:lambda_G}
	\end{equation}
	where $\lambda_G\to 0$ is the collapse length scale (see~\Cref{sb:scale}). The ground state $Q$ saturates the sharp GN inequality
	\begin{equation}
		\|u\|_{L^4}^4 \le \frac{2}{\|Q\|_{L^2}^2} \|u\|_{L^2}^2 \|\nabla u\|_{L^2}^2,
	\end{equation}
	and its $L^2$ norm determines the collapse threshold
	\begin{equation}
		G_* = \|Q\|_{L^2}^2 = 2\pi\int_0^\infty Q(r)^2 r\,\mathrm{d}r.
		\label{eq:GN}
	\end{equation}
	Solving \eqref{eq:elliptic_Q} numerically gives $Q(0)=2.2062$ and $G_*=11.7009$. The dichotomy between global existence and blow-up is well understood (see \cite{michaeli1983, BDN2023, WW2013}):
	\begin{itemize}
		\item \textbf{Global existence ($G > -G_*$)}:
		\Cref{eq:gp} admits global solutions for all $H^{1}$ initial data $\psi_0$ with $\|\psi_0\|_{L^2} = 1$.
		\item \textbf{Blow-up ($G < -G_*$)}:
		There exist initial data $\psi_0 \in H^{1} \cap L^2(|x|^2 {\rm d}x)$ such that the solution blows up in finite time, i.e. $\lim_{t \to T_{\max}} \|\psi\|_{H^{1}} = \infty$ for some $T_{\max} < \infty$.
	\end{itemize}
	Near criticality, as $G\to -G_*$ in the attractive regime $G<0$, the condensate is well approximated by the normalized rescaled profile $q=Q/\|Q\|_{L^2}$, and rigorous results confirm that the GP model captures the many-body dynamics up to blow-up time \cite{BDN2023}.
	
	\section{Numerical methods and results }
	
	\subsection{Local density approximation}
	
	To investigate the ground state of a 2D BEC with $G>0$ under near-critical rotation
	($\Omega \to 1$), we adopt a variational formulation restricted to the
	LLL. The condensate wave function is expanded in the orthonormal
	basis
	\begin{equation}
		\phi_k(z) = \frac{1}{\sqrt{\pi k!}} z^k e^{-|z|^2/2}, \quad k=0,1,\dots,N^{\mathrm{basis}}-1,
	\end{equation}
	with $z=x+\mathrm{i}y$; we write $\psi(z) = \sum_k c_k \phi_k(z)$, and the GP energy functional in the LLL subspace is minimized under
	the normalization constraint. The coefficients $c_k$ are treated as complex variational
	parameters, optimized by constrained minimization \cite{DB2020,BC2014}. An initial guess containing a hexagonal ring of vortices is projected onto the LLL, ensuring efficient convergence to vortex-lattice states closely resembling the multivortex clusters found in \cite{SAPS1998}.
	
	The resulting minimizers exhibit the characteristic features of rapidly rotating condensates.
	The macroscopic profile follows the TF envelope, while the microscopic structure
	reveals a triangular Abrikosov vortex lattice \cite{AD2001,MCWD2000}. Close to the trap center,
	the lattice is nearly perfect, whereas mild distortions appear near the boundary due to finite-size
	effects. The computed vortex density and interaction energy are consistent with the Abrikosov
	parameter $e^{\mathrm{Ab}}(1)$ \cite{NR2022}, confirming that the LLL variational
	framework accurately captures the interplay between rotation, interactions, and vortex formation.

\begin{figure}[t!]
   \begin{minipage}[t]{0.49\textwidth}
     \centering
     \includegraphics[width=\linewidth]{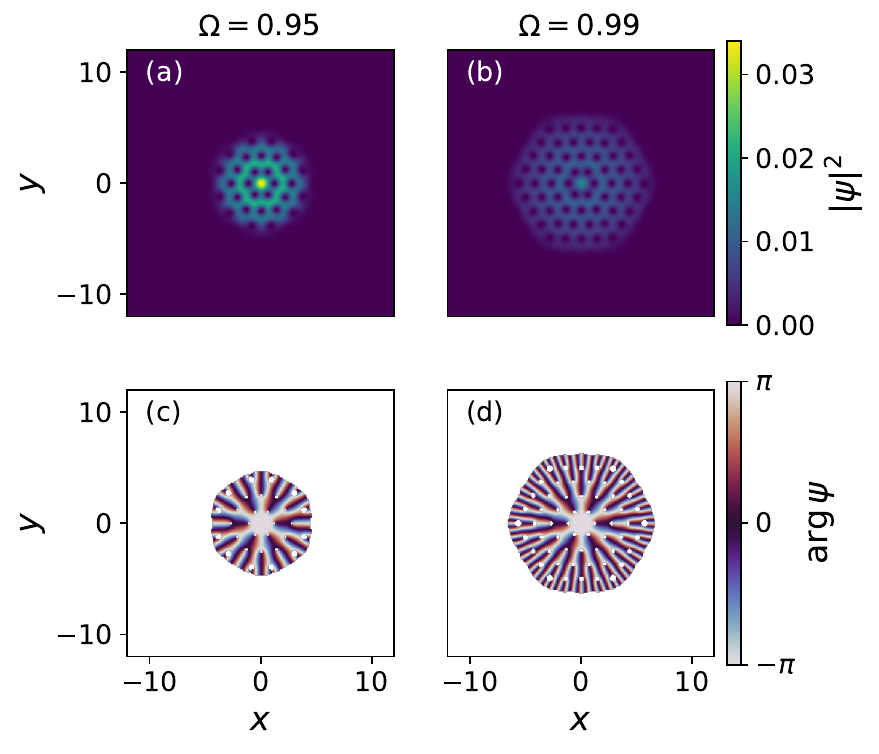}
     \caption{Density and phase of the condensate wave function $|\psi_\Omega|^2$ for $G=200$ in the weak effective trap $V^{\rm eff}_\Omega$ (see \eqref{eq:Veff}). Results are shown for rotation frequencies $\Omega=0.95$ and $\Omega=0.99$.}\label{fig: mean-field}
   \end{minipage}
   \begin{minipage}[t]{0.49\textwidth}
     \centering
     \includegraphics[width=\linewidth]{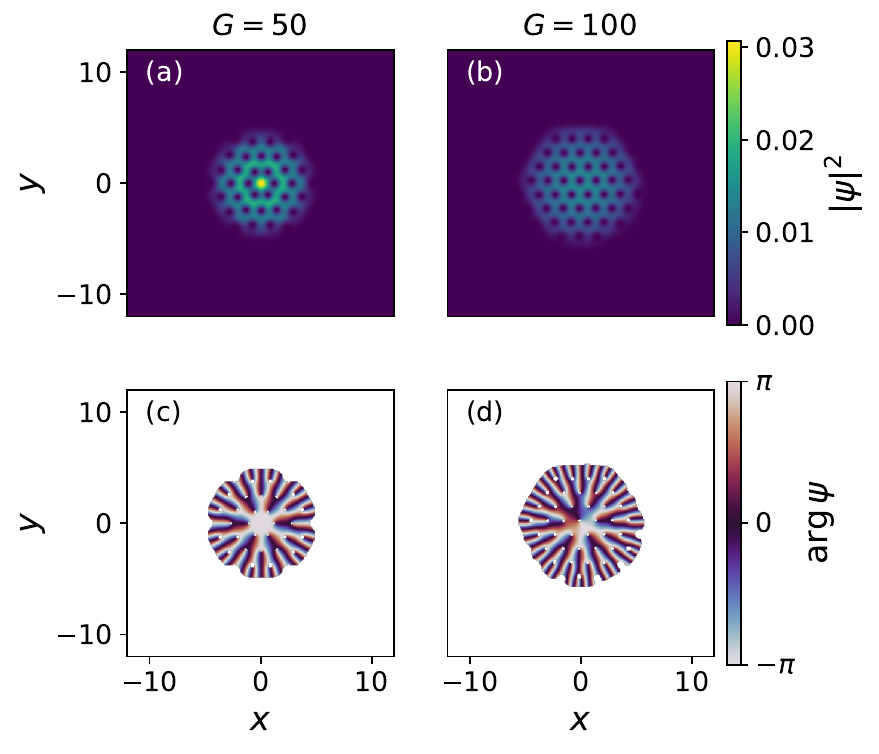}
     \caption{Density of the condensate wave function $|\psi_G|^2$ for repulsive interactions in the weak effective trap $V^{\rm eff}_{\Omega=0.99}$ (see \eqref{eq:Veff}). The vortex pattern and corresponding energy distributions are shown for interaction strengths $G=50$ and $G=100$.}
\label{fig:mean-field-G}
   \end{minipage}
\end{figure}


	\Cref{fig: mean-field} illustrates the formation of a vortex lattice closely resembling the Abrikosov structure observed in experiments \cite{MCWD2000}. The corresponding GP energy functional in the LLL is expressed in \eqref{eq:E_gp}, proportional to $G$, accounting for the interaction energy. As $\Omega$ increases, the interaction contribution becomes dominant, resulting in a flatter density distribution and a higher vortex density, consistent with energy minimization under the normalization constraint and its connection to the Abrikosov problem \cite{NR2022}.

	
	Then, we analyze the system at fixed rotational frequency. In \Cref{fig:mean-field-G}, for weak interactions, only a few vortices appear near the center, resulting in a sparse configuration. At intermediate coupling, the vortex array expands radially and develops into a nearly triangular Abrikosov lattice. For sufficiently strong interactions, the condensate radius grows markedly, and the vortex density converges to the Abrikosov constant, in agreement with the TF–LLL prediction \cite{NR2022} and observed in recent studies of BECs with spin–orbit coupling \cite{Banger2023}.
	
	The monotonic increase in vortex number with interaction strength underscores the dual role of repulsion: it suppresses density fluctuations while simultaneously driving spatial expansion, thereby facilitating vortex nucleation. Physically, stronger interactions render the condensate more incompressible, enabling it to accommodate a larger vortex lattice before energy minimization favors giant-vortex states or fragmentation at ultra-high, further reinforcing the analogy between rotating BECs and superconducting systems \cite{AB2006,tin-lun2001,AS2007}.

	We now specialize the TF--LLL functional, corresponding to a rotation harmonic trap $V^{\mathrm{eff}}_{\Omega}(x)$.
	The TF energy associated to the effective TF functional \eqref{eq:TF_LLL} with the minimizer has the inverted parabolic form \eqref{eq:p_TF}, supported in the disk $R_{\Omega,G}^{\mathrm{TF}} = \sqrt{\lambda_{\Omega,G}^{\mathrm{TF}}}$.
	Then, by substituting \eqref{eq:p_TF} into \eqref{eq:TF_LLL} and integrating explicitly, the Abrikosov constant can be expressed directly in terms of the minimal TF energy as
	\begin{equation}\label{eq:eAb-final}
		\boxed{\;
			e^{\mathrm{Ab}}(1)=\frac{9\pi}{2G(1-\Omega^2)}\;
			\left(\mathcal{E}^{\mathrm{\rm TF} }_{\Omega,G}\left[\rho^{\mathrm{\rm TF}}_{\Omega, G}\right]\right)^{2}\;}.
	\end{equation}
	
	The number of vortices is quantified by computing the curl of the phase gradient of the optimized wavefunction,
	\begin{equation}
		\text{curl} \theta = \partial_x \theta_y - \partial_y \theta_x,
		\label{eq:phase}
	\end{equation}
	where $ \theta_x $ and $ \theta_y $ are approximated using finite differences. Vortices are identified at points $ |\text{curl} \theta| \approx 2\pi $. 
	
	The Abrikosov parameter is quantified by the convention-free bulk ratio
	\begin{equation}
		e^{\rm sim} = \frac{\langle |\psi|^4 \rangle}{\langle |\psi|^2 \rangle^2},
		\label{eq:avg_density_vortex}
	\end{equation}
	evaluated over the central cells of the lattice, where $\langle f(x) \rangle \equiv S^{-1}\int_S f(x)\,\mathrm{d}x$ denotes the spatial average over a central region of area $S$. This numerical estimate is compared to the theoretical value $e^{\mathrm{Ab}}(1) \approx 1.1596$~\cite{AD2001}. \Cref{eq:avg_density_vortex} is the finite-volume counterpart of the theoretical definition, since, under the normalization $\int_S |\psi(x)|^2\mathrm{d}x = 1$, it reduces to $e^{\rm sim} = S\int_S |\psi(x)|^4\mathrm{d}x$, matching the integrand in~\eqref{eq:eAb_def}. In practice, however, finite grid resolution underresolves vortex cores, causing $e^{\rm sim}$ to underestimate the true value; we therefore use the energy-inversion estimate~\eqref{eq:eAb-final} as our primary benchmark.
	
	\begin{figure}[H]
		\centering
		\includegraphics[width=0.5\columnwidth]{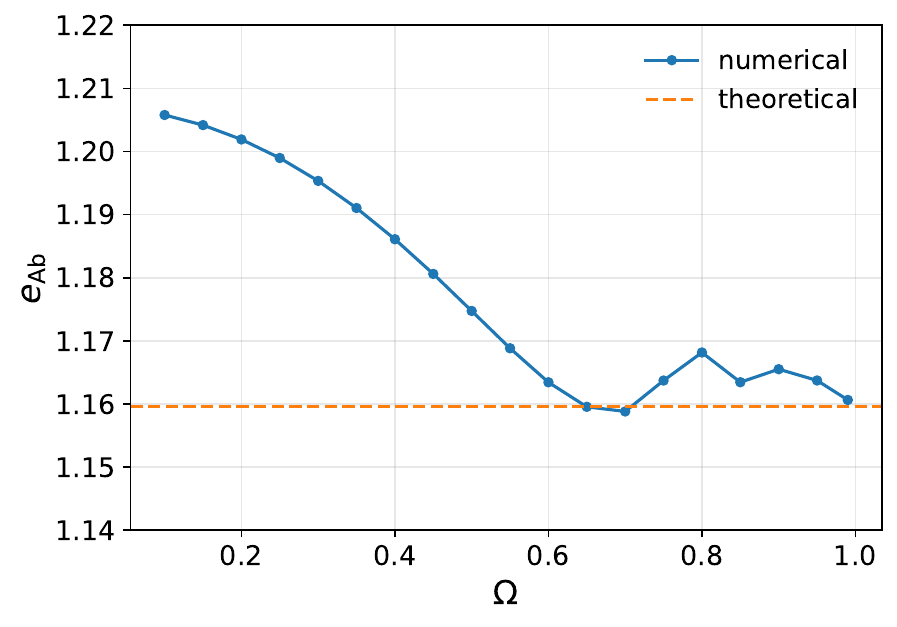}
		\caption{Abrikosov constant versus $\Omega$ at $G=200$: the numerical energy-inversion estimate $e^{\mathrm{Ab}}(1)$ of \eqref{eq:eAb-final} approaches the theoretical value $e^{\mathrm{Ab}}(1)=1.1596$ as $\Omega\to1$.}
		\label{fig:e_ab}
	\end{figure}
	
	In the TF--LLL regime, the Abrikosov constant follows from the minimal energy through the inversion \eqref{eq:eAb-final}: feeding the converged TF energy back into \eqref{eq:eAb-final} yields an estimate that decreases toward $e^{\rm Ab}(1)\approx1.1596$ and settles within about one percent of it for $\Omega\gtrsim0.6$ (see \Cref{fig:e_ab}). At slower rotation it sits above $1.1596$, where the kinetic energy is no longer negligible and the TF reduction underlying \eqref{eq:eAb-final} loses accuracy. The direct bulk ratio $e^{\rm sim}$ \eqref{eq:avg_density_vortex} measures the same constant, but on a finite grid it underresolves the vortex cores and so underestimates $\langle|\psi|^4\rangle$; we therefore use the energy inversion as the cleaner estimator here.
	
\begin{figure}[H]
		\centering
		\includegraphics[width=0.5\columnwidth]{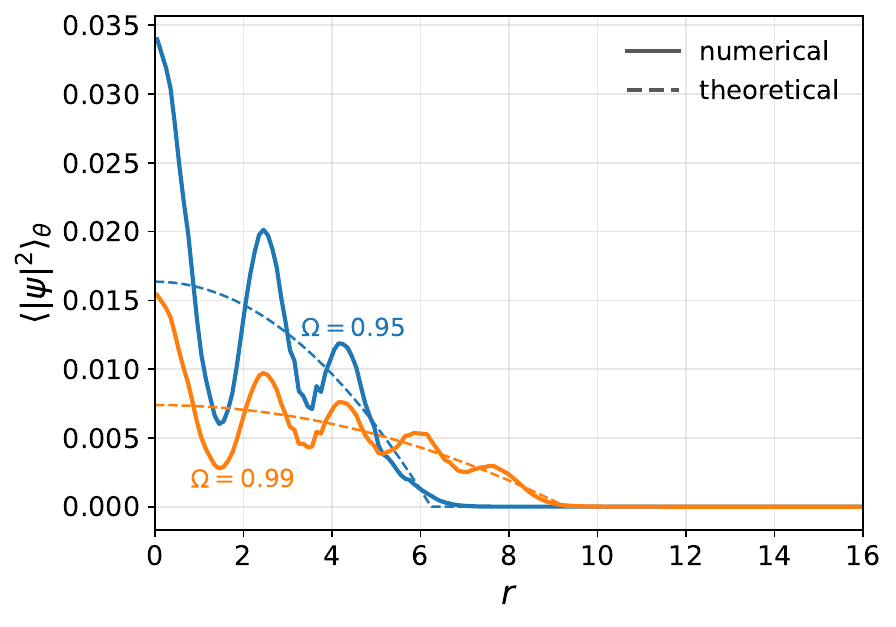}
		\caption{Rotationally averaged density: numerical (solid) and the theoretical TF profile \eqref{eq:p_TF} (dashed) for $G = 200$ (repulsive interaction), at $\Omega = 0.95$ and $0.99$.}
		\label{fig:Compare_TF}
	\end{figure}

	In order to enable a meaningful comparison with the condensate number density, we smooth out the rapid oscillations induced by the vortex lattice and contrast the rotationally averaged profile with the theoretical TF prediction (\Cref{fig:Compare_TF}). The agreement is closest at large $\Omega$, particularly regarding the TF radius, and the root-mean-square (RMS) error \cite{williamh2007}:
	\begin{equation}
		\mathrm{RMS} = \sqrt{\frac{1}{N} \sum_{i=1}^{N} \left[\rho^{\mathrm{sim}}(x_i) - \rho^{\mathrm{\rm TF}}(x_i)\right]^2},
		\label{eq:RMS}
	\end{equation}
	remains of order $10^{-2}$--$10^{-3}$ across a broad range of rotation frequencies $\Omega$, demonstrating that the optimized vortex–lattice solutions reproduce the theoretical TF profile in the strongly nonlinear regime. At smaller $\Omega$ the TF approximation is less accurate and the RMS error grows accordingly; we do not claim uniform agreement across all $\Omega$. In addition, the optimization procedure exhibits high computational efficiency, with nearly constant iteration counts and moderate execution times across all cases, confirming its suitability for large-scale parameter explorations.
	
	\subsection{Collapse in mean-field limit}
	For attractive interactions ($G<0$) confined in a harmonic trap and subject to rapid rotation, the condensate wavefunction $\psi(x,t)\in\mathbb{C}$ evolves according to the GP equation \eqref{eq:gp}, where $V^{\rm eff}_{\Omega}(x)$ denotes the external confinement given by \eqref{eq:Veff} and
	\[
	L_{z} = -\mathrm{i}\left(x\partial_y - y\partial_x\right).
	\]
	The normalization
	\begin{equation}
		\int_{[-L,L]^2} |\psi(x,t)|^2   {\rm d}x  = 1
		\label{eq:norm}
	\end{equation}
	is enforced at every step to ensure conservation of the total particle number.
	Then, we perform imaginary-time propagation $t\mapsto -\mathrm{i}\tau$ (gradient flow with discrete normalization \cite{BC2014,DGPS1999}), integrated by a time-splitting spectral scheme. One step $\psi^n\mapsto\psi^{n+1}$ reads
	\begin{equation}
		\begin{aligned}
			\tilde\psi &= e^{-\Delta\tau\left(\frac12|x|^2+\frac12 G|\psi^n|^2\right)}\,\psi^n, \\
			\hat\psi   &= e^{\Delta\tau\,\Omega L_z}\,\tilde\psi, \\
			\bar\psi   &= \mathcal{F}^{-1}\!\big[e^{-\frac12\Delta\tau\,k^2}\,\mathcal{F}\hat\psi\big], \\
			\psi^{n+1} &= \bar\psi/\|\bar\psi\|_{L^2},
		\end{aligned}
		\label{eq:gradient_flow}
	\end{equation}
	which exponentiates the trap and interaction exactly in real space and the kinetic term exactly in Fourier space. The rotation $-\Omega L_z$ is applied through its exact spectral derivative,
	\begin{equation}
		L_z\psi=-\mathrm{i}\,\mathcal{F}^{-1}\!\big[\,x\,(\mathrm{i}k_y\hat\psi)-y\,(\mathrm{i}k_x\hat\psi)\,\big],
		\label{eq:Lz_spectral}
	\end{equation}
	exponentiated to second order in $\Delta\tau\,\Omega$; unlike a real-space rigid rotation, no interpolation enters, so interpolation error is removed at the source. The final renormalization fixes the chemical potential, and the flow lowers the rotating-frame energy monotonically. Halving the step lowers the energy error at second order and the field error at close to second order (\Cref{tab:testA}); mass is conserved to machine precision.
	
	The computational domain is the square \( [-L,L]^2 \) discretized uniformly on an \(N\times N\) grid. All quantities are expressed in trap units. Convergence in the grid size $N$ is reported in \Cref{tab:testB}. Two of the attractive computations approach the collapse threshold too closely for a Cartesian grid, where grid anisotropy seeds an azimuthal symmetry breaking that drives a spurious collapse; for the charge-$m$ giant vortex near its threshold (\Cref{sb:giant}) we therefore relax the exact $\mathrm{e}^{\mathrm{i}m\theta}$ radial reduction in imaginary time, with a Crank--Nicolson kinetic step and an energy-monotonicity guard, and reconstruct the 2D field for display.

	\begin{figure}[H]
		\centering
		\includegraphics[width=\columnwidth]{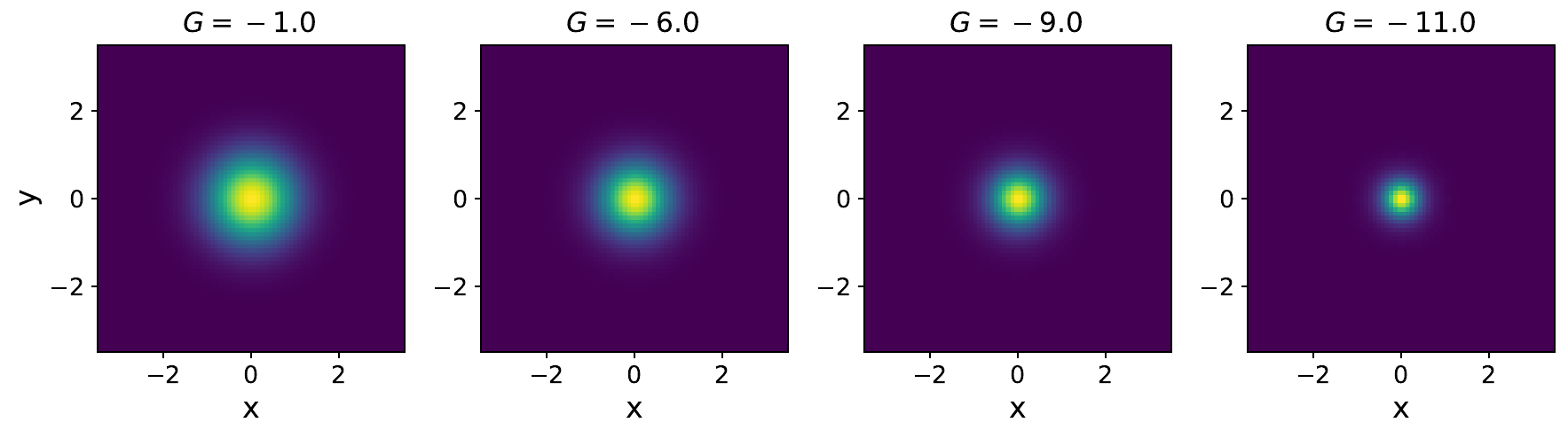}
		\caption{Attractive ground-state density $|\psi|^2$ at $\Omega=0.9$ for $G=-1,-6,-9,-11$, obtained by the imaginary-time flow \eqref{eq:gradient_flow}. The cloud is nodeless and contracts toward the center as $G\to -G_*$ with $G<0$; no vortex lattice forms at any of these couplings.}
		\label{fig:attractive}
	\end{figure}
		
	The outcome is qualitatively different from the repulsive case: the rotating ground state is a single nodeless lump, carrying no vortices at any coupling or rotation frequency (\Cref{fig:attractive}). Across $\Omega\in[0,0.99]$ the intrinsic energy is unchanged to the digits reported and the angular momentum remains negligible, growing only from $\sim10^{-18}$ at rest to $5\times10^{-6}$ at $\Omega=0.99$ (\Cref{tab:omega_attractive}). Only the effective-trap contribution $\tfrac{1-\Omega^2}{2}\langle|x|^2\rangle$ changes appreciably through its explicit $(1-\Omega^2)$ prefactor; here $\langle|x|^2\rangle$, defined in \eqref{eq:r}, is the mean-square radius. Rotation does not nucleate vortices here. The energetic reason is simple: with $G<0$ the interaction $\tfrac{G}{4}\int_{\mathbb{R}^2}|\psi(x)|^4\,\mathrm{d}x$ is lowered by concentrating density, exactly what the density node of a vortex core opposes. Unlike a repulsive condensate, the attractive ground state therefore carries essentially no angular momentum and does not support quantized circulation.

	When the attraction strength increases, the nonlinear term drives radial contraction of the condensate, leading to density accumulation at the center. From an energetic perspective, stronger attraction disfavors the maintenance of extended low-density structures, making states with enhanced central density more favorable. This mechanism is clearly reflected in the simulations: the central density rises and the mean-square radius decreases, where
	\begin{equation}
		\langle |x|^2 \rangle := \int_{\mathbb{R}^2} |x|^2 |\psi(x)|^2  \, \mathrm{d}x.
		\label{eq:r}
	\end{equation}
	As $G\to -G_*$ with $G<0$, the attractive interaction dominates over both quantum pressure and trapping, ultimately driving the condensate to collapse. In simulations, this regime manifests as a rapid growth of the peak density $|\psi(r)|_{\rm max}^2$, a strong negative shift of the interaction energy, and a marked reduction of $\langle |x|^2 \rangle$: the mean-square radius falls from $0.996$ at $G=-0.1$ to $0.268$ at $G=-11$, while the peak density rises from $0.32$ to $1.64$ (\Cref{tab:G_attractive}). The results in \Cref{fig:attractive} are in full agreement with theoretical predictions for 2D attractive condensates governed by the GP equation \cite{RHBE1995,GFT2001,UL1998}.
	
\begin{figure}[H]
		\centering
		\includegraphics[width=0.5\columnwidth]{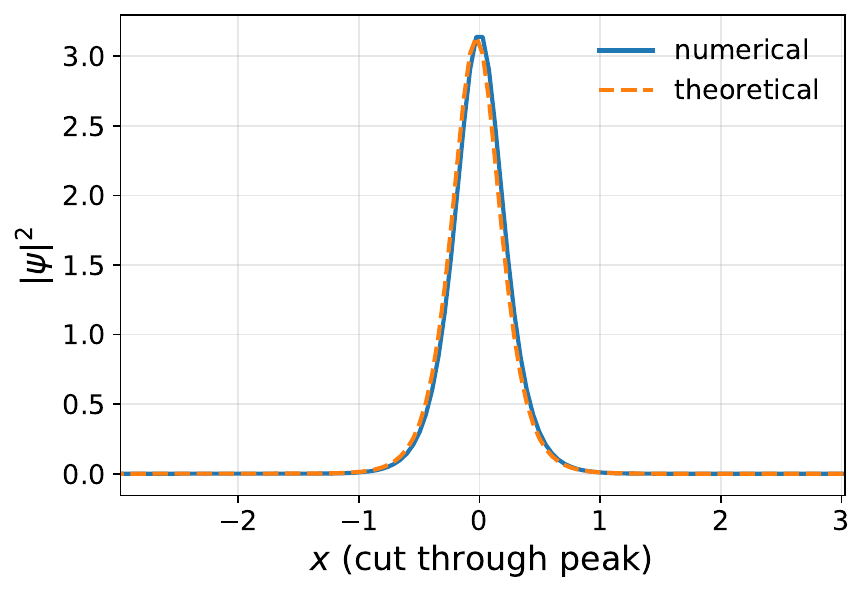}
		\caption{Comparison between the numerical profile (solid) and the theoretical rescaled Townes profile \eqref{eq:lambda_G} (dashed) along a one-dimensional $x$-cut through the peak density, close to threshold at $G=-11.5$, $\Omega=0.9$, $\lambda_G=0.364$.}
		\label{fig:Compare_Attractive}
	\end{figure}

	Collapse persists for all rotation frequencies, including the near-critical regime $\Omega \to 1$, where the condensate converges to the universal $Q$-profile modulo magnetic translations \cite{NR2025,DNR2023}. Numerical benchmarks confirm converged energies and peak densities across $\Omega$ (see~\Cref{tab:G_attractive,tab:omega_attractive}). These results demonstrate that collapse is ultimately dictated by the balance of kinetic and attractive contributions, largely independent of trap or rotation. Comparison with the rescaled elliptic ground state \eqref{eq:elliptic_Q}, using the dilation factor $\lambda_G$ from \eqref{eq:lambda_G}, shows close agreement near threshold ($G=-11.5$, $\lambda_G=0.364$), confirming that the simulations capture the correct collapse profile (\Cref{fig:Compare_Attractive}).
	
	\subsection{Vortex expulsion and giant vortices}
	\label{sb:giant}
	The vortex arrays reported for attractive condensates in an earlier version of this study were imprinted; they are not ground states. Within the LLL, the energy \eqref{eq:E_gp} is minimized for $G<0$ by \emph{maximizing} $\int_{\mathbb{R}^2}|\psi(x)|^4\,\mathrm{d}x$, whereas the triangular lattice minimizes it (the Abrikosov problem \cite{ABN2006}); the lattice is thus the highest-energy LLL configuration under attraction and cannot be a local minimum.

	\begin{figure}[H]
		\centering
		\includegraphics[width=\columnwidth]{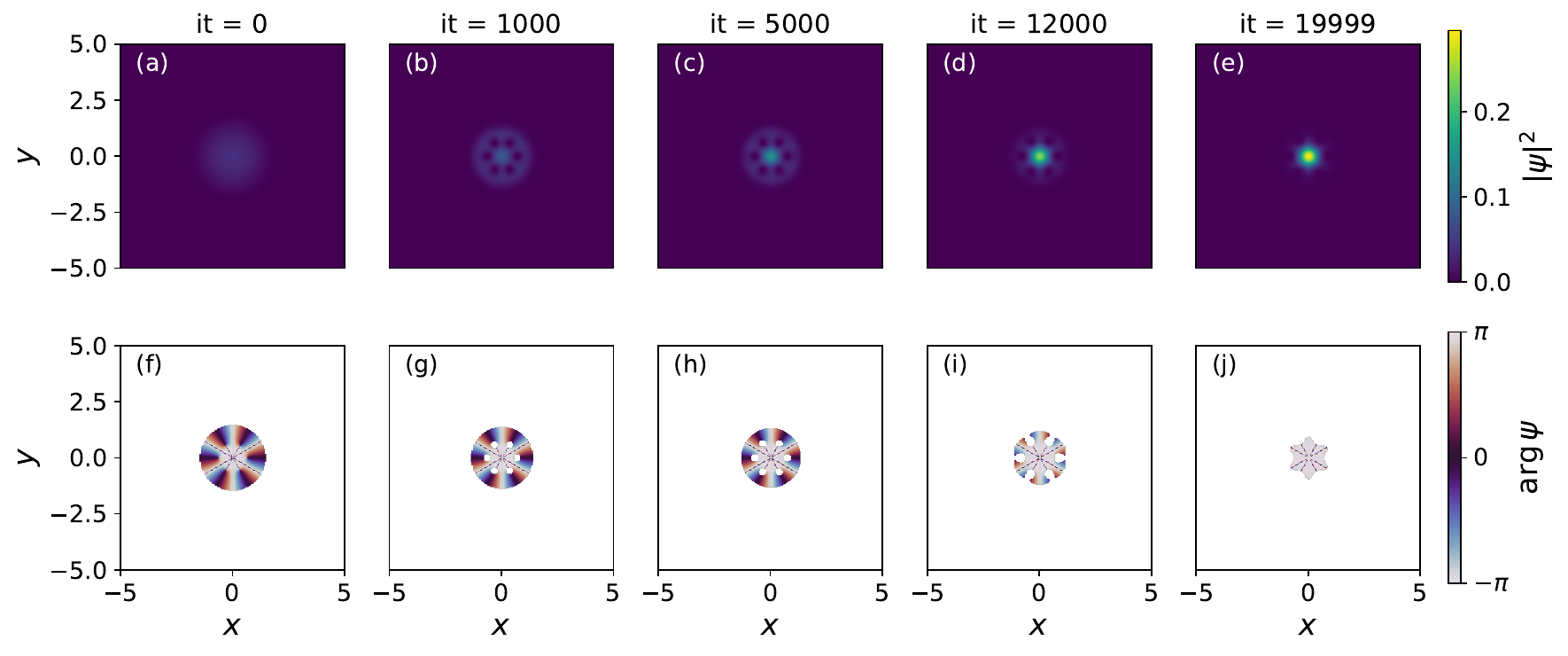}
		\caption{Imaginary-time relaxation of a tight six-vortex ring at $G=-0.05$, $\Omega=0.97$ (top: $|\psi|^2$; bottom: $\arg\psi$). The vortices drift outward and the density concentrates at the centre; $L_z$ falls from $4.5$ toward $0$ and the relaxed state is vortex free.}
		\label{fig:relax}
	\end{figure}
	
	Imprinting phase singularities raises the total kinetic energy through the flow term \cite{FS2001, PS2016}:
	\begin{equation}
		E^{\rm flow} = \frac{1}{2}\!\int_{\mathbb{R}^2} \rho(x)|\nabla S(x)|^2\, \mathrm{d}x,
	\end{equation}
	which represents the kinetic energy of the superfluid current associated with the phase gradient $\mathbf{v}=\nabla S$. Imaginary-time relaxation makes the resulting instability explicit: starting from a tight ring of six singly-quantized vortices at $G=-0.05$, $\Omega=0.97$, the angular momentum decreases monotonically from $L_z\simeq4.5$ toward zero as the vortices drift outward and the density concentrates at the centre (\Cref{fig:relax}); the relaxed state is the nodeless lump. The cloud expels its vorticity rather than supporting it, even at $\Omega=0.97$.

	\begin{figure}[H]
		\centering
		\includegraphics[width=0.5\columnwidth]{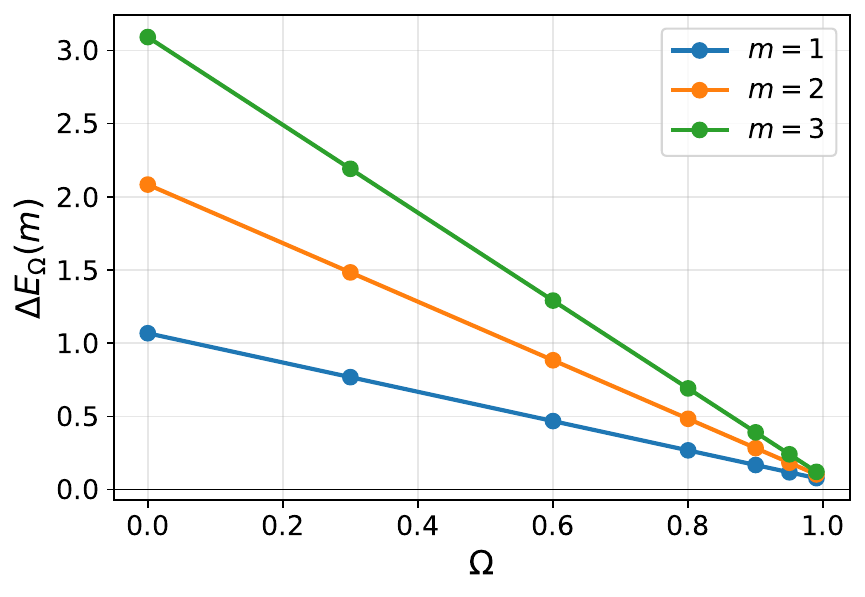}
		\caption{Rotating-frame excitation $\Delta E_\Omega(m)$ of the charge-$m$ giant vortex above the nodeless ground state at $G=-3$. It is positive for all $\Omega<1$, vanishing only as $\Omega\to1$.}
		\label{fig:giant_exc}
	\end{figure}
		
	Whether any vortex state survives is answered by restricting the flow to a fixed angular-momentum sector. Seeding $\psi_0=\mathrm{e}^{\mathrm{i}m\theta}f(r)$ pins a charge-$m$ core at the origin; because the rotating-frame functional is rotationally covariant, the flow stays in that sector and relaxes to a stationary \emph{giant vortex} of charge $m$, with $L_z=m$ to numerical precision. Writing $\psi_m$ for this relaxed charge-$m$ state and $\mathcal E(m):=\mathcal{E}^{\rm GP}_{\Omega,G}[\psi_m]$ for its rotating-frame GP energy \eqref{eq:gp_energy_rbec} (with $\psi_0$ the nodeless ground state), the excitation $\Delta E_\Omega(m)=\mathcal E(m)-\mathcal E(0)$ stays positive for every $\Omega<1$, vanishing only as $\Omega\to1$ (\Cref{fig:giant_exc}), so rotation never makes a vortex favorable below the centrifugal limit.

\begin{figure}[H]
		\centering
		\includegraphics[width=\columnwidth]{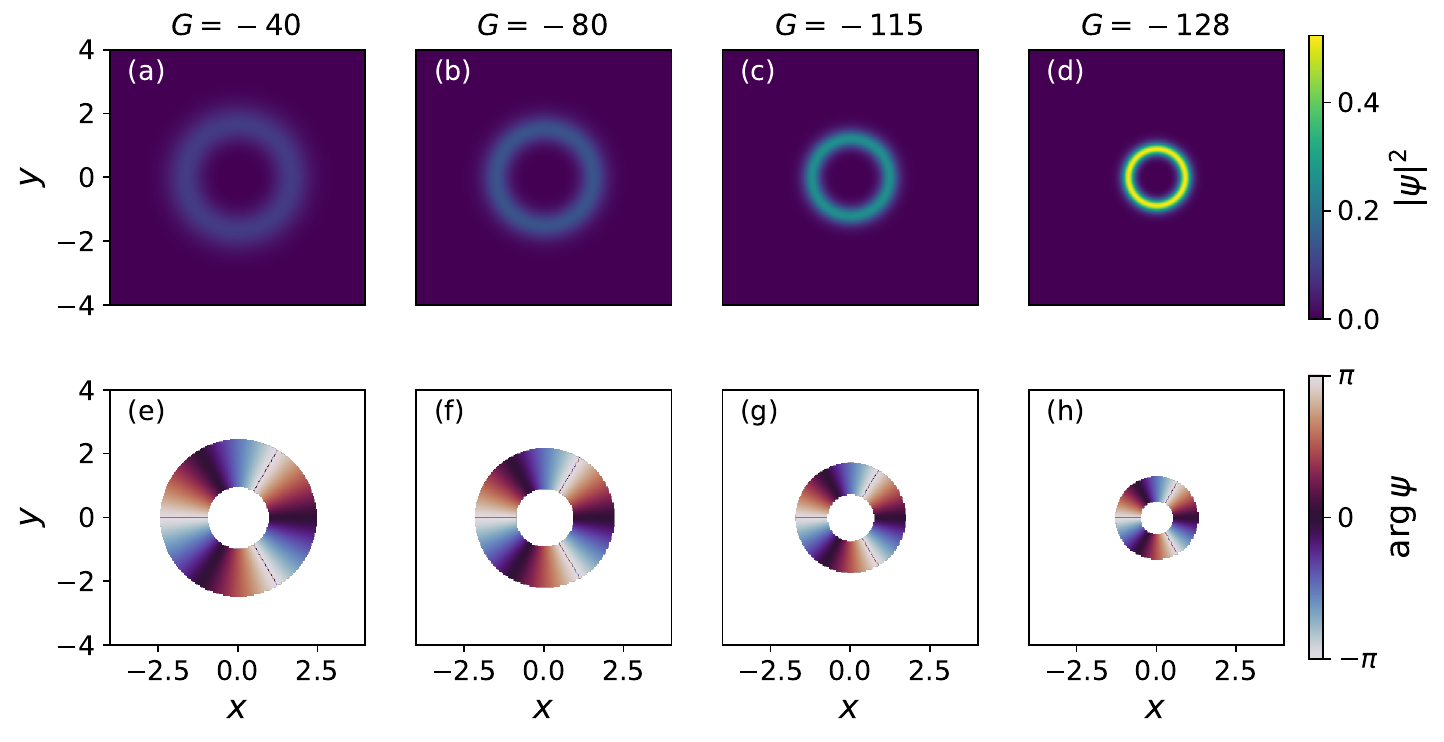}
		\caption{Stationary charge-$3$ giant vortex at $\Omega=0.9$, driven toward its collapse threshold $G=-\|Q_3\|_{L^2}^2\simeq-132.4$ for $G=-40,-80,-115,-128$
			(top: $|\psi|^2$; bottom: $\arg\psi$), computed in the exact $\mathrm{e}^{3\mathrm{i}\theta}$ radial reduction. The ring tightens as $G\to -\|Q_3\|_{L^2}^2$, contracting onto the rescaled charge-3 soliton $Q_3$; $L_z=3$ throughout.}
		\label{fig:giant}
	\end{figure}

\begin{figure}[H]
		\centering
		\includegraphics[width=\columnwidth]{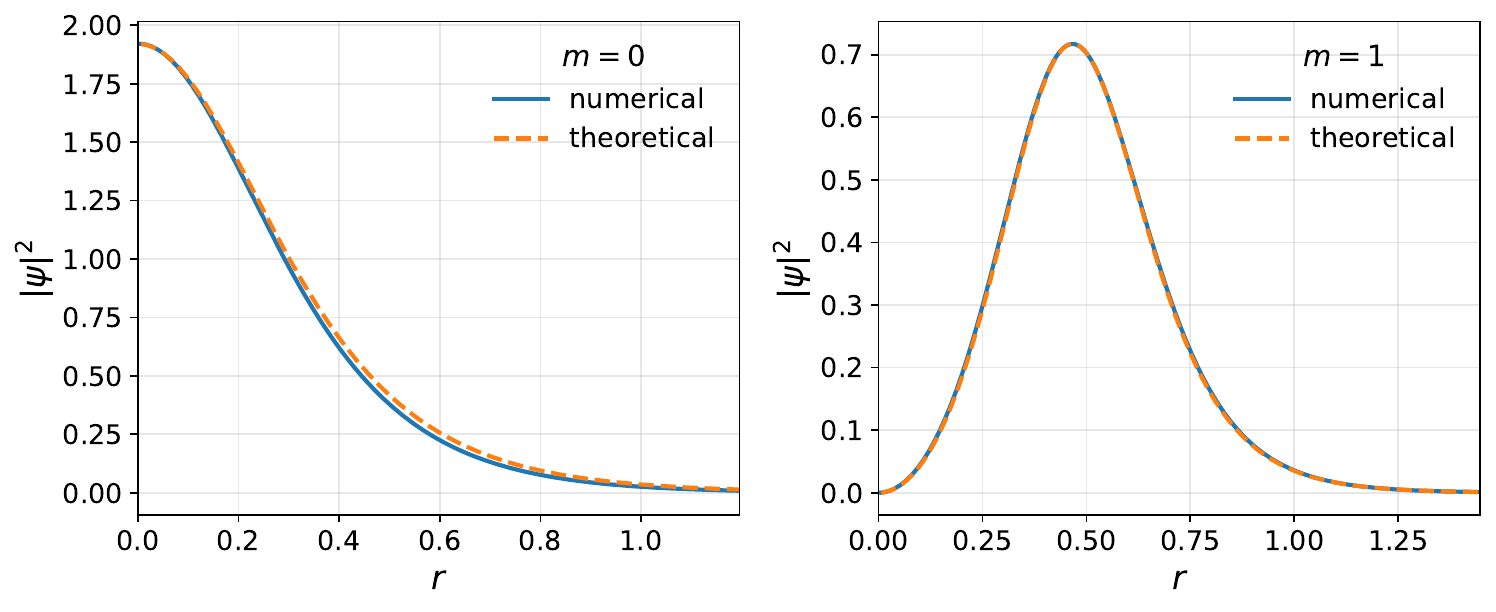}
		\caption{Radial density of the giant vortex driven near its own threshold ($G=-0.97\,\|Q_m\|^2$, $\Omega=0.9$; numerical, solid) against the rescaled soliton density $R_m(r/\lambda_m)^2$ (theoretical, dashed), with fitted rescaling length $\lambda_m$: the lump ($m=0$, threshold $11.70$) and the charge-1 ring ($m=1$, threshold $48.3$).}
		\label{fig:Qcompare_m}
	\end{figure}

	These rings are trapped, rotating realizations of the \emph{vortex (Townes) soliton} of the focusing equation: near its threshold the charge-$m$ core converges, under the rescaling in \Cref{app:collapse}, to the charge-$m$ ground state $Q_m=\mathrm{e}^{\mathrm{i}m\theta}R_m(r)$ of \eqref{eq:elliptic_Q}, the vortex generalization of the Townes profile \cite{kruglov1985,desyatnikov2005}. Solving \eqref{eq:elliptic_Q} by the same shooting that fixes $Q_0$ gives the charge-$m$ critical masses reported in \Cref{tab:Qm}; these masses are each well above $\|Q_0\|_{L^2}^2=G_*\simeq11.70$; an equivariant GN inequality makes $G_*(m)=\|Q_m\|_{L^2}^2$ the collapse threshold of the charge-$m$ sector (\Cref{app:collapse}). The giant vortex is therefore \emph{not} collapse-free: within its symmetry class it is trap-stabilized for $-\|Q_m\|_{L^2}^2<G<0$ and collapses onto $Q_m$ as $G\to -\|Q_m\|_{L^2}^2$. We follow this contraction with the symmetry-exact radial solver: \Cref{fig:giant} drives the charge-3 ring to $G=-128$, close to $-\|Q_3\|_{L^2}^2\simeq-132.4$, where it has tightened onto the rescaled $Q_3$, and \Cref{fig:Qcompare_m} overlays the relaxed $m=0$ and $m=1$ cores on their rescaled solitons. Once $G<-\|Q_0\|_{L^2}^2$, the symmetry no longer protects the ring in the full space: it can collapse through the azimuthal (modulational) instability \cite{firth1997,desyatnikov2005}; the trap, the rotation, and the fixed-$L_z$ sector are what suppress this channel here.
	
	\section{Discussion}
	\subsection{Vortex counting: TF vs. ODE predictions}
	
	We conducted numerical simulations for systems with repulsive and attractive interactions, varying the rotation frequency $\Omega$. The number of vortices was determined from the phase winding, using the coarse-grained formula derived in \Cref{vt:ct}.

	\begin{figure}[H]
		\centering
		\includegraphics[width=0.5\columnwidth]{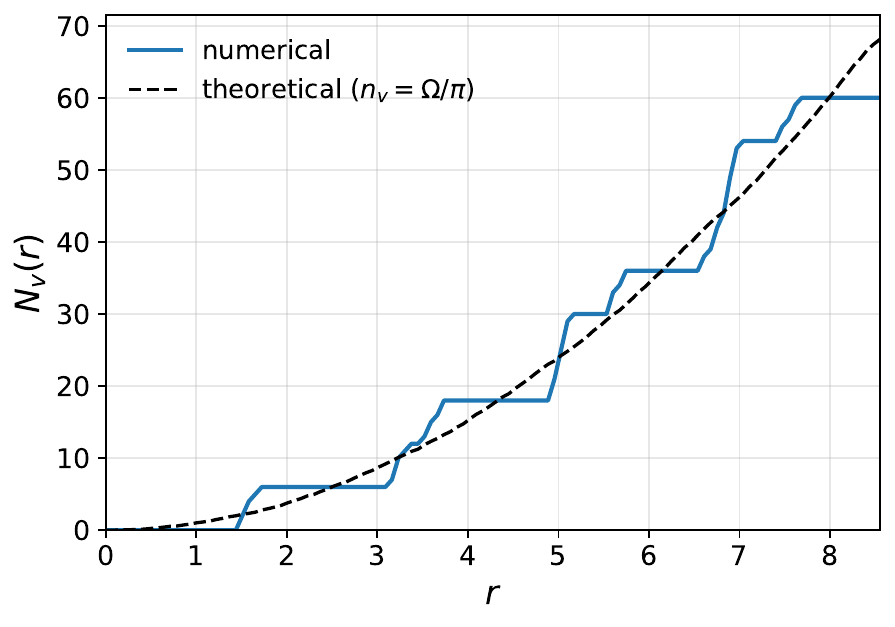}
		\caption{Cumulative vortex count $N_v(r)$ of the repulsive lattice (numerical, solid) against the theoretical uniform Feynman law $n_v=\Omega/\pi$ cut off by the finite cloud (dashed), at $G=200$, $\Omega=0.99$. For attractive interactions the count is zero at every $\Omega$.}
		\label{fig:count_vortex}
	\end{figure}
		
	For repulsive interactions, the cumulative count $N_v(r)$ grows as the rigid-body law $N_v\simeq\Omega r^2$ through the bulk and saturates at the cloud edge, tracking the uniform Feynman density $n_v=\Omega/\pi$ cut off by the finite condensate (\Cref{fig:count_vortex}). The quantitative agreement between the computed count and the theoretical scaling law \eqref{eq:vortex-TF} confirms that the counting functional captures the interplay between centrifugal expansion and mean-field repulsion in determining vortex multiplicity; at fixed $\Omega$ the total count also increases monotonically with $G$, reflecting the larger condensate radius (\Cref{fig:mean-field-G}).
	
	In contrast, for attractive interactions the measured count is zero at every rotation frequency, consistent with the negligible angular momentum of \Cref{tab:omega_attractive}: the enhanced compressibility of the condensate counteracts centrifugal spreading and disfavors vortex nucleation altogether \cite{NR2025}. The concomitant radial contraction, quantified by the second spatial moment \eqref{eq:r}, confirms that the two interaction signs realize opposing mechanisms—centrifugal expansion versus collapse.

	\subsection{Abrikosov constants of superfluidity and superconductivity}
	A central question concerns the comparison between Abrikosov constants in type-II superconductors and rapidly rotating BECs. In the Ginzburg--Landau (GL) framework, one restricts attention, near the second critical field $H_{c2}$, to periodic configurations on a fundamental lattice cell $Q_{L}$ and introduces the Abrikosov functional (under the normalization $\int_{Q_{L}} |u(x)|^{2}\,\mathrm{d}x = 1$) by \cite{AS2007}
	\begin{equation}
		\label{eq:fAb_def}
		f^{\mathrm{Ab}} (1)
		= \lim_{L\to\infty}
		\inf_{\substack{u \in \mathcal{LLL}_L \\ \|u\|_{L^2(Q_L)}=1}}
		\frac{1}{|Q_L|}
		\int_{Q_L} \big(|u(x)|^4 - 2|u(x)|^2\big) \, \mathrm{d}x,
	\end{equation}
	where $Q_L$ denotes a square of side length $L$ satisfying the flux quantization
	$|Q_L|=L^2\in 2\pi\mathbb{N}$, and the admissible set
	$\mathcal{LLL}_{L}$ is the corresponding finite-dimensional restriction of the LLL to $Q_L$,
	that is, functions of the form \eqref{eq:LLL} with holomorphic $f(z)$ compatible with the
	magnetic periodicity of the vortex lattice \cite{AB2006,ABN2006_vortex}.
	
	In the GP framework for rapidly rotating BECs, the 2D energy reduces to the LLL functional defined by the Abrikosov constant~\eqref{eq:eAb_def}. Both constants are minimized by the triangular lattice.
	
	Rigorous analysis implies the inequality $e^{\mathrm{Ab}}(1) f^{\mathrm{Ab}}(1) \leq -1$, with the conjectured sharp identity $e^{\mathrm{Ab}}(1) f^{\mathrm{Ab}}(1) = -1$ in the thermodynamic limit \cite{AS2007}. A single number therefore fixes both constants. Our repulsive minimizers pin that number through the energy inversion \eqref{eq:eAb-final}, $e^{\rm Ab}(1)\approx1.1596$ (\Cref{fig:e_ab}); the bulk ratio \eqref{eq:avg_density_vortex} measures the same constant directly but underestimates it on a finite grid, where the vortex cores are underresolved. Under the conjectured identity this fixes the companion constant $f^{\rm Ab}(1)\approx-0.86$.
	
	\subsection{Collapse ratio in the attractive regime}
	\label{sb:scale}
	To quantify condensate contraction under attractive interactions, we use the unit-mass profile $q=Q/\|Q\|_{L^2}$ and analyze $u_G(x)=\lambda_G^{-1}q(\lambda_G^{-1}x)$.
	Near the critical coupling $G\to -G_*$ in the attractive regime one finds \cite{GS2014,DNR2023,LNR2018}
	\begin{equation}
		\lambda_G := \frac{1}{\ell} \sim (G+G_*)^{1/4},
	\end{equation}
	so that the effective radius vanishes as in \eqref{eq:r}, while the collapse rate diverges set by $\ell$.
	A detailed derivation is provided in~\Cref{app:collapse}. The same construction extends to each angular-momentum sector: the charge-$m$ vortex soliton $Q_m$ obeys an equivariant GN inequality with threshold $G_*(m)=\|Q_m\|_{L^2}^2$, rising as $11.70,\,48.3,\,89.8,\,132.4$ for $m=0,\dots,3$ (\Cref{app:collapse}), so that the giant vortex of \Cref{sb:giant} collapses onto the rescaled $Q_m$ at its own threshold.
	
	\subsection{Effectiveness of numerical methods for repulsive interaction}
	
	We benchmarked four representative solvers for rotating BECs:
	\begin{enumerate}[label=(\roman*)]
		\item the trust-region constrained optimizer (\textsc{trust-constr}) from \textsc{SciPy} \cite{VGO2020},
		\item imaginary-time evolution (ITE) projected onto the LLL basis \cite{FS2001},
		\item real-space split-step ITE \cite{BC2014, DGPS1999},
		\item the Riemannian Conjugate Gradient (RCG) method \cite{WW2013, bart201, sato2022}.
	\end{enumerate}
	
\begin{figure}[t!]
   \begin{minipage}[t]{0.49\textwidth}
     \centering
     \includegraphics[width=\linewidth]{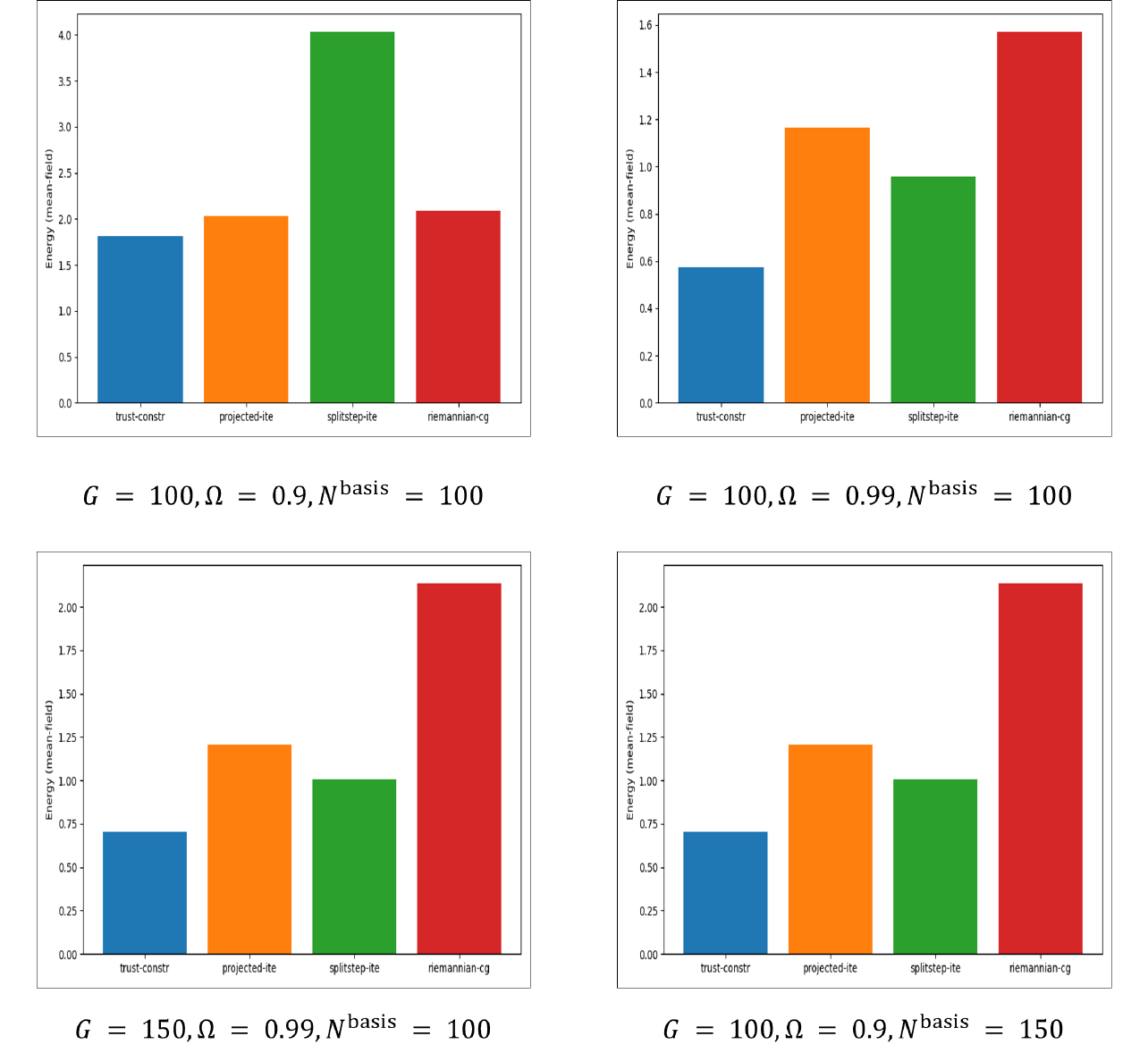}
		\caption{Comparison of ground-state energies obtained by different numerical methods. Each panel corresponds to a parameter set $(G,\Omega,N^{\textrm{basis}},N_x)$.}
		\label{fig:energy_comparison}
   \end{minipage}
   \begin{minipage}[t]{0.49\textwidth}
     \centering
     \includegraphics[width=\linewidth]{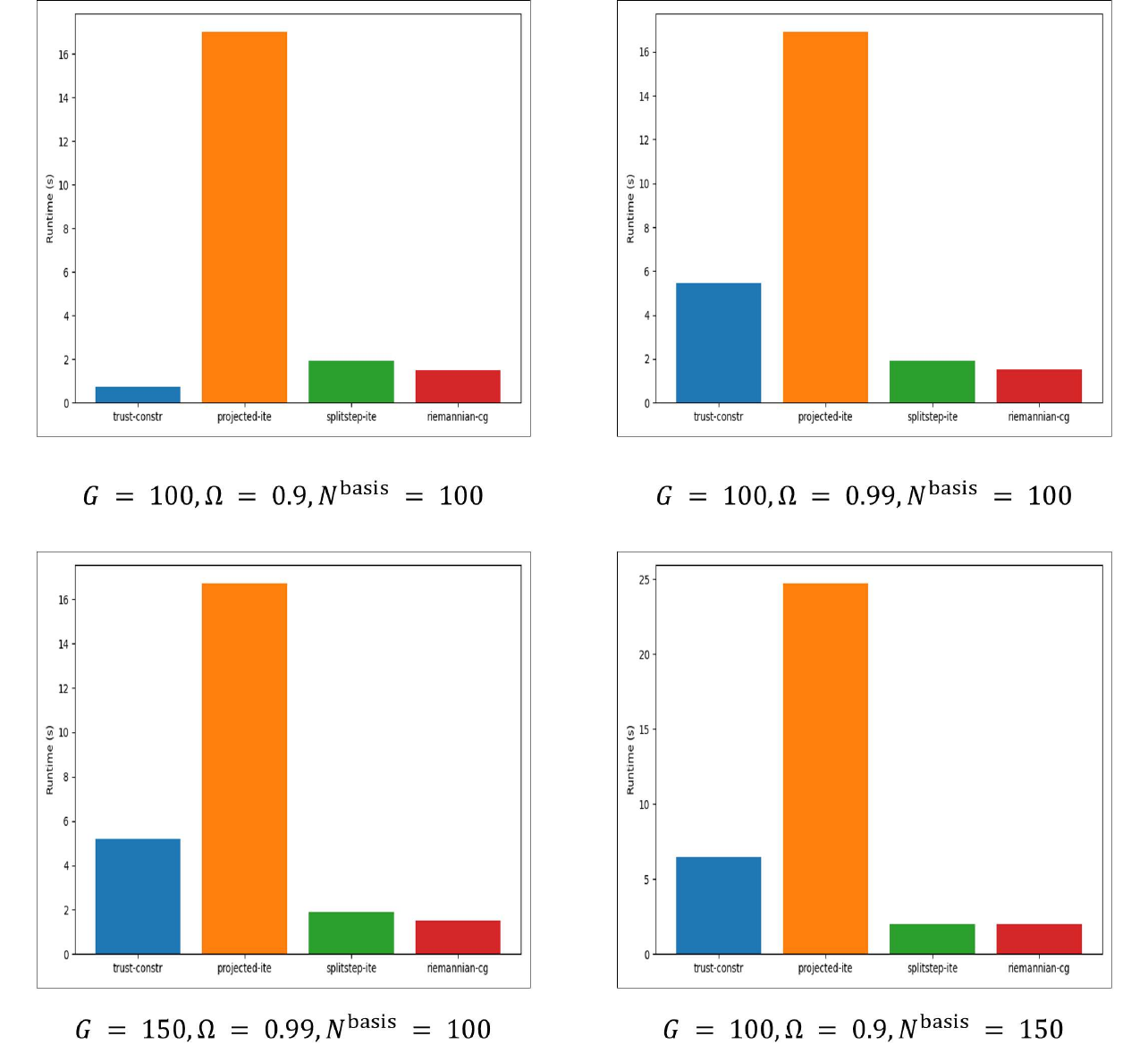}
     \caption{Wall-clock times required by different methods. Each panel corresponds to a parameter set $(G,\Omega,N^{\textrm{basis}},N_x)$.}
\label{fig:time_comparison}
   \end{minipage}
\end{figure}

	
	Ground-state energies were computed for representative cases ($G=100,150$, $\Omega=0.9,0.99$), see \Cref{fig:energy_comparison}. Among all methods, \textsc{trust-constr} consistently produced the lowest energies with nearly perfect overlap ($\approx 1.0$) to the reference, demonstrating superior accuracy. Projected-ITE and RCG showed partial convergence, while split-step ITE generated spurious vortex proliferation ($\sim 10^2$), see \Cref{tab:vortex_count}, confirming its unreliability near $\Omega\to 1$ without gauge projection.

	Wall-clock timings (\Cref{fig:time_comparison}) show RCG as the fastest ($\sim 1.8$ s), followed by split-step ITE ($\sim 2$ s).
	\textsc{trust-constr} required 4–6 s but remained the most robust.
	Projected-ITE was the slowest ($\sim 20$ s) due to repeated basis projections.
	Thus, variational optimizers are preferable for high-accuracy ground states, while Riemannian methods are suitable for rapid prototyping.
	
	Fidelity was further assessed by vortex counting (\Cref{tab:vortex_count}).
	Only \textsc{trust-constr} reproduced the Abrikosov lattice vortex number \cite{CWG2001}, while split-step ITE produced unphysical excess vortices, and the other methods yielded intermediate results.
	
	Overall, these results confirm that optimization-based solvers with explicit constraints provide the most reliable description of rotating condensates, especially near the fast-rotation limit.
	
	\subsection{Effectiveness of numerical methods for attractive interaction}
	\label{sb:attractive_interaction}
	
	For $G<0$, we performed two convergence tests:
	\begin{enumerate}[label=(\roman*)]
		\item temporal accuracy (Test A)
		\item spatial resolution consistency (Test B)
	\end{enumerate}
	The results are reported in \Cref{tab:testA,tab:testB}.
	
	Test A shows that halving $\Delta t$ reduces the energy error by a factor of four, confirming second-order temporal convergence, with the field error decaying at close to second order.
	Mass conservation is preserved at machine precision, and energy drift scales with $\Delta t^2$.
	
	Test B demonstrates spatial convergence: the total energy is converged to $\sim10^{-7}$ and the peak density to $\sim10^{-2}$ over the tested grids.
	
	Taken together, the scheme is both \emph{accurate} and \emph{robust} for attractive condensates, even close to collapse thresholds \cite{LNR2018, NR2025, thi2020}.
	In particular:
	\begin{itemize}
		\item clear second-order convergence in time,
		\item exact mass conservation,
		\item controlled energy drift,
		\item grid-converged energies and observables.
	\end{itemize}
	
	\subsection{Validity: noise and finite-temperature effects}
	The whole analysis above is deterministic mean-field at zero temperature. Thermal depletion, dissipation, and quantum and thermal fluctuations are not included, and they bear directly on the regimes considered here: fluctuations seed and pin vortices, broaden the collapse threshold, and would modify the vortex expulsion of \Cref{sb:giant}; noise need not be purely destructive, and can act constructively to sustain or lock nonlinear excitations such as solitons and breathers \cite{desantis2023,guarcello2025}. A stochastic or finite-temperature treatment \cite{noise2002,noise2004,noise2008,noise2016a,noise2016b,weiler2008,blakie2008,gardiner2002} would be needed to follow these effects quantitatively; we expect them to round the sharp $\Omega$-independence and threshold reported here without changing the qualitative separation between the two interaction signs.
	
	\section{Conclusions}

	In summary, we have studied vortex patterns of a two-dimensional condensate at the almost critical rotation speed, treating each interaction sign with the method it requires. The LLL variational treatment reproduces the triangular Abrikosov lattice, the TF envelope, and the Abrikosov constant $e^{\rm Ab}(1)\approx1.1596$ through the energy inversion \eqref{eq:eAb-final}. The real-space treatment shows that the attractive ground state carries no vortex lattice at any rotation: the focusing nonlinearity favors a nodeless, contracting cloud that collapses as $G\to -G_*$, with $G_*=\|Q\|_{L^2}^2\simeq11.70$, essentially independently of $\Omega$. An imprinted lattice is expelled under relaxation, and the stationary vortex states we find are giant vortices—trapped vortex Townes solitons—each carrying its own collapse threshold $G=-\|Q_m\|_{L^2}^2$ ($\|Q_m\|_{L^2}^2=11.70,\,48.3,\,89.8,\,132.4$ for $m=0,\,1,\,2,\,3$), fixed analytically by an equivariant Gagliardo--Nirenberg inequality and confirmed numerically up to the critical region. Whether the sharp rotation-independence survives at finite temperature, where fluctuations compete with the kinetic boundary layer, is open.
	
	We end with a remark on the strictly critical speed $\Omega=1$. In the attractive case the analysis could be extended to $\Omega=1$ itself: although the effective trap \eqref{eq:Veff} vanishes there, a GP ground state still exists for $-G_*<G<0$, bound by the focusing nonlinearity of the GP functional alone. This is a mathematical peculiarity of the nonlinear mean-field description rather than a feature of the underlying physics: at the many-body level the ground state ceases to exist at maximal rotation, where the centrifugal force exactly cancels the trap and nothing confines the particles. Since the present work treats the repulsive and attractive regimes side by side, and the repulsive lattice requires $\Omega<1$, we have restricted for simplicity to the almost critical regime throughout.

	\appendix
	
	\section{Comparison with experiment for attractive interactions}
	\label{s:at_exp}
	
	We map our dimensionless parameters to the collapse experiment of
	Bradley \cite{bradley1997} on $^7$Li, with $a_s=-1.45(4)$~nm, $m=7$~u, $\omega_{x,y}\simeq2\pi\times151$~Hz, $\omega_z\simeq2\pi\times132$~Hz, and observed critical number $650\lesssim N_c^{\rm(exp)}\lesssim1300$. With $a_z=\sqrt{\hbar/(m\omega_z)}$, $\kappa=\sqrt{8\pi}\,a_s/a_z$, and $G=2\kappa N$, collapse at $G=-G_*$ gives $N_c^{\rm sim}=G_*/(2|\kappa|)$. The stated $\omega_z$ fixes $a_z\simeq3.31\,\mu$m and $\kappa\simeq-2.20\times10^{-3}$, so that with $G_*=11.70$,
	\begin{equation}
		N_c^{\rm sim}\simeq2.66\times10^{3}.
	\end{equation}
	This is the same order as the measured window but a factor of two above it. The discrepancy is expected: the trap is nearly spherical ($\omega_z\simeq\omega_{x,y}$), so the gas is three-dimensional rather than quasi-two-dimensional, and the strict 2D reduction over-counts the atoms needed for collapse. We therefore read the comparison as order-of-magnitude consistency, not a quantitative match (\Cref{tab:sim-exp}).
	
	\section{Derivation of the collapse scaling law}
	\label{app:collapse}
	
	Let $Q$ denote the Townes profile of \eqref{eq:elliptic_Q}, with $\|Q\|_{L^2}^2=G_*$, and set $q=Q/\|Q\|_{L^2}$. We start from the unit-mass rescaled profile
	\begin{equation}
		u_\ell(x) = \ell q(\ell x), \quad \|q\|^{2}_{L^2(\mathbb{R}^2)}=1.
	\end{equation}
	Substituting this ansatz into the GP energy functional \eqref{eq:gp_energy_rbec}
	\begin{equation}
		\mathcal{E}^{\rm GP}_{\Omega,G}[u] = \frac{1}{2}\int_{\mathbb{R}^2} |\nabla u(x)|^2 \,\mathrm{d}x + \frac{1}{2}\int_{\mathbb{R}^2} |x|^2 |u(x)|^2 \,\mathrm{d}x + \frac{G}{4}\int_{\mathbb{R}^2} |u(x)|^4 \,\mathrm{d}x
	\end{equation}
	and denoting,
	\[
	A = \int_{\mathbb{R}^2} |\nabla q(x)|^2 \,\mathrm{d}x,\quad
	B = \int_{\mathbb{R}^2} |x|^2 q(x)^2 \,\mathrm{d}x,\quad
	C = \int_{\mathbb{R}^2} q(x)^4 \,\mathrm{d}x,
	\]
	(the rotation term vanishes for the radial profile), for $-G_*<G<0$ with $G\to -G_*$, we obtain
	\begin{equation}
		\mathcal{E}^{\rm GP}_{\Omega,G}[u]
		= \frac{1}{2}\Big(A + \frac{G}{2}C\Big)\ell^2
		+ \frac{B}{2}\ell^{-2}.
	\end{equation}
	The critical coupling $G_*$ is defined by
	\(
	A - \frac{G_*}{2}C = 0,
	\)
	i.e. $G_* = 2A/C$. The Pohozaev identities for $Q$ imply $A=1$ and $C=2/\|Q\|_{L^2}^2$ for $q$, whence $G_*=\|Q\|_{L^2}^2=11.70$, the GN threshold \eqref{eq:GN}.
	For $-G_*<G<0$ with $G\to -G_*$ one has
	\begin{equation}
		\mathcal{E}^{\rm GP}_{\Omega,G}[u]
		= \frac{C}{4}(G+G_*)\ell^2 + \frac{B}{2}\ell^{-2}.
	\end{equation}
	Minimizing the above in $\ell>0$ yields
	$$
	\lambda_G = \frac{1}{\ell} \sim (G+G_*)^{1/4}.
	$$
	This proves that as $G\to -G_*$, the condensate collapses with diverging contraction rate $\ell$ and vanishing length scale $\lambda_G$, consistent with \eqref{eq:r}.
	
	The construction extends to every angular-momentum sector. For equivariant states $\psi=\mathrm{e}^{\mathrm{i}m\theta}u(r)$ the kinetic energy carries the centrifugal term $\int_{\mathbb{R}^2}\big(|u'(r)|^2+m^2u(r)^2/r^2\big)\,\mathrm{d}r$, and the sharp equivariant GN inequality reads
	\begin{equation}
		\|\psi\|_{L^4}^4\le\frac{2}{\|Q_m\|_{L^2}^2}\,\|\psi\|_{L^2}^2\,\|\nabla\psi\|_{L^2}^2 ,
		\label{eq:GN_m}
	\end{equation}
	saturated by the charge-$m$ vortex soliton $Q_m=\mathrm{e}^{\mathrm{i}m\theta}R_m(r)$. Its radial profile satisfies
	\begin{equation}
		-R_m''-\frac{1}{r}R_m'+\frac{m^2}{r^2}R_m+R_m-R_m^3=0.
	\end{equation}
	Since the $L^2$-critical scaling in two dimensions is blind to $m$, $Q_m$ obeys the same Pohozaev identities as $Q_0$, and the charge-$m$ collapse threshold is the soliton mass $G_*(m)=\|Q_m\|_{L^2}^2$, raised above the lump value by the centrifugal barrier. Solving this radial equation by shooting on the $r^m$ amplitude, and checking the Pohozaev ratios, reproduces them to five digits and yields the thresholds of \Cref{tab:Qm}. The protection is sectorial: \eqref{eq:GN_m} restricts to equivariant data, so in the full space the sharp constant reverts to $2/\|Q_0\|^2$, and a charge-$m$ ring with mass above $\|Q_0\|^2$ can still collapse by shedding its symmetry.
	
	\section{Vortex-counting formula}
	\label{vt:ct}
	We briefly derive the coarse-grained formula used to estimate the number of vortices enclosed within a disk of radius $r$. Let $\psi:\mathbb{R}^2\to\mathbb{C}$ be the condensate wave function with density $\rho(x)=|\psi(x)|^2$.
	The vorticity of the phase $\varphi=\arg\psi$ satisfies, in the sense of
	distributions \cite{feynman1955, fetter2009,DNR2023}
	\begin{equation}
		\nabla\times\nabla\varphi = 2\pi\sum_j q_j\delta_{x_j},
	\end{equation}
	where $q_j\in\mathbb{Z}$ denotes the winding number of the zero $x_j$.
	Upon coarse-graining at a scale larger than the vortex cores, the repulsive lattice mimics rigid-body rotation, $\langle\nabla\varphi\rangle\approx\Omega\,x^\perp$, so that
	\[
	\nabla\times\langle\nabla\varphi\rangle = 2\Omega = 2\pi n_v,
	\]
	which is Feynman's rule: a uniform areal vortex density $n_v=\Omega/\pi$ inside the condensate, independent of the local density. The cumulative vortex number inside radius $r$ is then the uniform density times the enclosed condensate area,
	\begin{equation}\label{eq:vortex-TF}
		\boxed{
			N_{v, G > 0}(r) = \frac{\Omega}{\pi}\,A_{\rm cl}(r),\qquad
			A_{\rm cl}(r)=\int_{|x|<r}\theta\big(\rho(x)-\rho_c\big)\,\mathrm{d}x ,}
	\end{equation}
	with $\rho_c$ a small density cutoff. This grows as $\Omega r^2$ in the bulk and saturates at the cloud edge; it is the expression used in the main text for comparing numerical simulations with the theoretical prediction (\Cref{fig:count_vortex}).
	
	\paragraph{Attractive interactions.}
	For $G<0$ the measured count is zero at every rotation frequency, consistent with $L_z\approx0$ in \Cref{tab:omega_attractive}: the ground state carries no circulation, so the coarse-grained relation above does not apply. Numerically we count phase-winding plaquettes inside the bulk (a radial cut for the lattice, a smoothed-density mask for the peaked attractive cloud).
	
	\section{Tables}
	\begin{table}[H]
		\caption{Charge-$m$ vortex Townes solitons $Q_m$ of \eqref{eq:elliptic_Q} by shooting: collapse threshold $G_*(m)=\|Q_m\|_{L^2}^2$, its ratio to the lump value, and the Pohozaev ratios $\int_{\mathbb{R}^2}|\nabla Q_m(x)|^2\,\mathrm{d}x/\|Q_m\|_{L^2}^2$ and $\int_{\mathbb{R}^2}|Q_m(x)|^4\,\mathrm{d}x/\|Q_m\|_{L^2}^2$, equal to $1$ and $2$ for every $m$.}
		\label{tab:Qm}
		\begin{tabular}{ccccc}
			\hline
			$m$ & $\|Q_m\|_{L^2}^2$ & $G_*(m)/G_*(0)$ & $\int_{\mathbb{R}^2}|\nabla Q_m(x)|^2\,\mathrm{d}x/\|Q_m\|_{L^2}^2$ & $\int_{\mathbb{R}^2}|Q_m(x)|^4\,\mathrm{d}x/\|Q_m\|_{L^2}^2$ \\
			\hline
			0 & 11.70 & 1.00 & 1.00000 & 2.00000 \\
			1 & 48.30 & 4.13 & 1.00000 & 2.00000 \\
			2 & 89.75 & 7.67 & 1.00000 & 2.00000 \\
			3 & 132.42 & 11.32 & 1.00000 & 2.00000 \\
			\hline
		\end{tabular}
	\end{table}
	\begin{table}[H]
		\caption{\label{tab:vortex_count}%
			Vortex count comparison for $G=100$, $\Omega=0.99$,
			$N^{\rm basis}=100$, $N_x=192$.}
		\begin{tabular}{lcccc}
			\hline
			Method          & Energy & Vortices & Overlap & Time (s) \\
			\hline
			Trust-constr     & 0.572 & 92   & 1.00 & 5.4 \\
			Projected-ITE    & 1.165 & 47   & 0.34 & 16.9 \\
			Splitstep-ITE    & 0.958 & 129 & 0.00 & 1.9 \\
			Riemannian-CG    & 1.569 & 51   & 0.24 & 1.5 \\
			\hline
		\end{tabular}
	\end{table}
	
	\begin{table}[H]
		\caption{Collapse threshold in experiment \cite{bradley1997} and simulations.}
		\label{tab:sim-exp}
		\begin{tabular}{lcc}
			\hline
			Quantity & Experiment & Simulation \\
			\hline
			Trap frequencies &
			$\omega_{x,y}\!\simeq\!2\pi\times151$ Hz, & Used in mapping \\
			& $\omega_z\!\simeq\!2\pi\times132$ Hz & ($a_z=3.31\,\mu$m) \\
			Scattering length $a_s$ & $-1.45(4)$ nm & Input to $\kappa$ \\
			Critical $N_c$ & $650$--$1300$ & $\simeq2.66\times10^{3}$ \\
			Rotation $\Omega$ & $0$ & $0 \leq \Omega \leq 0.99$ \\
			\hline
		\end{tabular}
	\end{table}
	
	\begin{table}[H]
		\centering
		\caption{Temporal convergence test (Test A), $G=-6$, $\Omega=0.9$, $N=192$.
			Reported are the relative $L^2$ error of the field, the drift in the conserved mass,
			and the error in the total energy for different time step sizes $\Delta t$. Mass is conserved to machine precision; the energy error is second order in $\Delta t$ and the field error close to second order.}
		\label{tab:testA}
		\begin{tabular}{c c c c}
			\hline
			$\Delta t$ & rel $L^2$ error & mass drift & energy error \\
			\hline
			$2.0\times 10^{-1}$ & $9.96\times 10^{-2}$ & $0$ & $1.72\times 10^{-2}$ \\
			$1.0\times 10^{-1}$ & $5.38\times 10^{-2}$ & $2.2\times 10^{-16}$ & $5.36\times 10^{-3}$ \\
			$5.0\times 10^{-2}$ & $2.76\times 10^{-2}$ & $2.2\times 10^{-16}$ & $1.55\times 10^{-3}$ \\
			$2.5\times 10^{-2}$ & $1.33\times 10^{-2}$ & $0$ & $4.17\times 10^{-4}$ \\
			$1.25\times 10^{-2}$ & $5.79\times 10^{-3}$ & $0$ & $1.04\times 10^{-4}$ \\
			$6.25\times 10^{-3}$ & $1.94\times 10^{-3}$ & $0$ & $2.11\times 10^{-5}$ \\
			\hline
		\end{tabular}
	\end{table}
	
	\begin{table}[H]
		\centering
		\caption{Spatial convergence test (Test B), $G=-6$, $\Omega=0.9$, $\Delta t=2.5\times10^{-4}$.
			The energy is converged to $\sim10^{-7}$ and the peak density to $\sim10^{-2}$ over this range.}
		\label{tab:testB}
		\begin{tabular}{c c c c}
			\hline
			Grid $N$ & $E_{\rm tot}$ & $|\psi|^2_{\max}$ & $\langle |x|^2\rangle$ \\
			\hline
			128 & 0.7150102 & 0.48718 & 0.7160 \\
			256 & 0.7150102 & 0.49952 & 0.7160 \\
			384 & 0.7150102 & 0.50183 & 0.7160 \\
			\hline
		\end{tabular}
	\end{table}
	
	\begin{table}[H]
		\centering
		\caption{Attractive ground state versus $G$ at $\Omega=0.9$ ($N=256$, $L=12$). The cloud contracts ($\langle |x|^2\rangle\downarrow$, peak$\uparrow$) toward collapse; $L_z$ stays negligible, i.e. no vortices.}
		\label{tab:G_attractive}
		\begin{tabular}{cccccc}
			\hline
			$G$ & $E_{\rm tot}$ & $E_{\rm kin}$ & $E_{\rm int}$ & $\langle |x|^2\rangle$ & $|\psi|^2_{\max}$ \\
			\hline
			$-0.1$ & 0.9960 & 0.5019 & $-0.0040$ & 0.9963 & 0.3186 \\
			$-1.0$ & 0.9593 & 0.5213 & $-0.0417$ & 0.9596 & 0.3355 \\
			$-3.0$ & 0.8711 & 0.5754 & $-0.1399$ & 0.8715 & 0.3828 \\
			$-6.0$ & 0.7150 & 0.7090 & $-0.3519$ & 0.7159 & 0.4995 \\
			$-9.0$ & 0.5028 & 1.0420 & $-0.7912$ & 0.5051 & 0.7845 \\
			$-11.0$ & 0.2627 & 2.0970 & $-1.9680$ & 0.2676 & 1.6435 \\
			\hline
		\end{tabular}
	\end{table}
	
	\begin{table}[H]
		\centering
		\caption{Attractive ground state versus $\Omega$ at $G=-9$. The intrinsic state is rotation-independent: $E_{\rm tot}$ is constant and $L_z\approx0$; only the effective trap energy $E^{\rm eff}_{\rm trap}=\tfrac{1-\Omega^2}{2}\langle |x|^2\rangle$ varies.}
		\label{tab:omega_attractive}
		\begin{tabular}{cccc}
			\hline
			$\Omega$ & $E_{\rm tot}$ & $E^{\rm eff}_{\rm trap}$ & $|L_z|$ \\
			\hline
			0.00 & 0.5028 & 0.2526 & $2\times10^{-18}$ \\
			0.30 & 0.5028 & 0.2298 & $3\times10^{-18}$ \\
			0.50 & 0.5028 & 0.1894 & $1\times10^{-17}$ \\
			0.70 & 0.5028 & 0.1288 & $2\times10^{-13}$ \\
			0.90 & 0.5028 & 0.0480 & $3\times10^{-8}$ \\
			0.95 & 0.5028 & 0.0246 & $7\times10^{-7}$ \\
			0.99 & 0.5028 & 0.0050 & $5\times10^{-6}$ \\
			\hline
		\end{tabular}
	\end{table}
	
\medskip
\noindent{\bf Acknowledgement.} This research is funded by University of Science, VNU-HCM under grant number T2025-14.

\medskip
\noindent{\bf Data availability.} The data that support the findings of this study are available from the corresponding author upon reasonable request.

\medskip
\noindent{\bf Declarations.} The authors have no conflict of interest to disclose.

\bibliographystyle{siam}

\end{document}